\documentclass{article}
\usepackage{spconf,amsmath,graphicx}

\usepackage{amssymb}
\usepackage{booktabs}
\usepackage[numbers]{natbib}
\usepackage{subfigure}
\usepackage{soul}

\usepackage{enumitem} 
\newcommand{\hide}[1]{} %hide
\usepackage{xspace}
\newcommand*{\modelname}{{uaMix-MAE}\@\xspace}

\newcommand\scalemath[2]{\scalebox{#1}{\mbox{\ensuremath{\displaystyle #2}}}}

\newcommand*{\ie}{\emph{i.e.},\@\xspace}
\newcommand*{\etc}{\emph{etc.}\@\xspace}

\usepackage{colortbl}
\definecolor{mygraylite}{gray}{.94}
\definecolor{mygray}{gray}{.89}
\definecolor{darkergreen}{RGB}{21, 152, 56}
\definecolor{amber}{rgb}{1.0, 0.75, 0.0}
\definecolor{darkseagreen}{rgb}{0.56, 0.74, 0.56}
\definecolor{babyblueeyes}{rgb}{0.63, 0.79, 0.95}
\definecolor{burntsienna}{rgb}{0.91, 0.45, 0.32}
\definecolor{myviolet}{rgb}{0.57, 0.51, 0.9}
\definecolor{amethyst}{rgb}{0.6, 0.4, 0.8}
\usepackage{tabularx}
\usepackage{pifont}
\usepackage{arydshln}

\usepackage{pgfplots}
\usetikzlibrary{plotmarks}
\usepgfplotslibrary{fillbetween}
\pgfplotsset{compat=1.15}
\pgfplotsset{
    discard if not/.style 2 args={
        x filter/.code={
            \edef\tempa{\thisrow{#1}}
            \edef\tempb{#2}
            \ifx\tempa\tempb
            \else
                
            \fi
        }
    }
}

\usepackage[pagebackref,breaklinks,colorlinks]{hyperref}

\usepackage[scr=rsfs]{mathalpha}
\usepackage[capitalize]{cleveref}
\crefname{section}{Sec.}{Secs.}
\Crefname{section}{Section}{Sections}
\Crefname{table}{Table}{Tables}
\crefname{table}{Tab.}{Tabs.}
\title{\modelname: Efficient Tuning of Pretrained Audio Transformers with Unsupervised Audio Mixtures}
\name{\begin{tabular}{c}Afrina Tabassum$^{1, \star}$,Dung Tran$^{2}$, Trung Dang$^{2}$, Ismini Lourentzou$^{1,3}$, Kazuhito Koishida$^{2}$\end{tabular}
\thanks{$^{\star}$ Work done during an internship at Microsoft}
}
\address{$^{1}$ Virginia Tech, $^{2}$Applied Sciences Group, Microsoft Corporation \\$^{3}$University of Illinois at Urbana - Champaign}
\begin{document}
%\ninept
%
\maketitle
\begin{abstract}
Masked Autoencoders (MAEs) learn rich low-level representations from unlabeled data but require substantial labeled data to effectively adapt to downstream tasks. 
Conversely, Instance Discrimination (ID) emphasizes high-level semantics, offering a potential solution to alleviate annotation requirements in MAEs. Although combining these two approaches can address downstream tasks with limited labeled data, naively integrating ID into MAEs leads to extended training times and high computational costs.
To address this challenge, we introduce \modelname, an efficient ID tuning strategy that leverages unsupervised audio mixtures. Utilizing contrastive tuning, \modelname aligns the representations of pretrained MAEs, thereby facilitating effective adaptation to task-specific semantics. 
To optimize the model with small amounts of unlabeled data, we propose an audio mixing technique that manipulates audio samples in both input and virtual label spaces. 
Experiments in low/few-shot settings demonstrate that \modelname achieves $4-6\%$ accuracy improvements over various benchmarks when tuned with limited unlabeled data, such as AudioSet-20K. Code is available at \href{https://github.com/PLAN-Lab/uamix-MAE}{https://github.com/PLAN-Lab/uamix-MAE}  
\end{abstract}

\section{Introduction}
Self-supervised learning has attracted significant attention for its ability to learn meaningful representations from vast amounts of unlabeled data, mitigating the need for costly annotations. Besides significant advancements in computer vision~\cite{he2020momentum, chen2020simple, he2022masked} and natural language processing~\cite{brown2020language, touvron2023llama, kenton2019bert}, self-supervised learning has also recently demonstrated potential for various speech and audio understanding tasks~\cite{huang2022masked, Chen2022beats, baade2022mae, niizumi2021byol, saeed2021contrastive, wu2022wav2clip}. Two highly effective self-supervised techniques in speech and audio understanding are Masked Audio Modeling (MAM)~\cite{huang2022masked, Chen2022beats, baade2022mae, gong2021ast} exemplified by methods such as MAE~\cite{huang2022masked}, and Instance Discrimination (ID) ~\cite{niizumi2021byol, saeed2021contrastive, wu2022wav2clip, heggan2023mtslvr}. 

MAE~\cite{he2022masked} employs a pre-training task where audio inputs are partitioned into non-overlapping patches, and a subset of these patches is masked and reconstructed using Transformer architectures such as ViT~\cite{dosovitskiy2020image}. Training objectives include patch reconstruction loss~\cite{he2022masked, gong2021ast} and discrete label prediction~\cite{Chen2022beats}. However, MAE representations often lack semantic alignment (\ie alignment of representations to capture intra-class similarities) as the reconstruction loss predominantly focuses on low-level time-frequency features while overlooking high-level semantic features~\cite{bao2021beit, ramesh2021zero, epstein2019generalization}. As a result, they require significant amounts of labeled data for effective adaption to downstream tasks. To address this limitation, BEATs~\cite{Chen2022beats} trains an acoustic tokenizer alongside an audio self-supervised model iteratively, albeit at the cost of increased model complexity and training time, yielding only marginal improvements in downstream tasks and performing less optimally in low/few-shot scenarios.

In contrast, ID methods, such as contrastive learning (CL), semantically align representations of different augmentations of the same audio input~\cite{niizumi2021byol}. Specifically, CL brings multiple augmentations of the same example (positive samples) closer while pushing other examples (negative samples) farther apart by utilizing an instance classification pretext strategy~\cite{chen2020simple}. In the image domain, \citet{lehner2023contrastive} proposes to combine CL with Masked Image Modeling (MIM) to extract object-centric representations by disregarding background details, thereby alleviating substantial annotation requirements in downstream tasks. However, this approach requires large-scale unlabeled datasets~\cite{imagenet15russakovsky}, resulting in increased training time and computational overhead. Thus, combining ID and MAE to tackle downstream tasks with constrained labeled data remains challenging.

\noindent\textbf{Our contributions.} In this work, we introduce \textit{\modelname}, an efficient ID contrastive tuning strategy with \underline{u}nsupervised \underline{a}udio \underline{Mix}tures for pretrained \underline{MAE}s, which enables effective adaptation to downstream tasks, particularly in low/few-shot settings, while only requiring small amounts of unlabeled data for MAE model tuning. \modelname initializes a ViT encoder with model weights trained with MAM~\cite{huang2022masked} and tunes the model, using a contrastive objective, with unsupervised audio mixtures to semantically align representations of pretrained MAEs. Moreover, inspired by~\cite{shen2022mix, ViT2040}, we propose a mixture technique tailored for audio that manipulates both the input and virtual label spaces simultaneously. This encourages the model to learn more precise and smoother decision boundaries in the latent feature space while training with small amounts of unlabeled data. Experimental results on several benchmark datasets show that \modelname outperforms strong MAM baselines by $4-6\%$ in low/few-shot scenarios.
\section{Related Work}\label{sec:related_work}

\noindent \textbf{Masked Audio Modeling (MAM)} has been applied to various audio understanding~\cite{huang2022masked, Chen2022beats, chong2023masked, niizumi2022masked}, natural language processing~\cite{kenton2019bert}, and computer vision tasks~\cite{dosovitskiy2020image}. As masked audio models, such as AudioMAE~\cite{he2022masked}, MaskSpec~\cite{chong2023masked}, MSM-MAE~\cite{niizumi2022masked}, BEATs~\cite{Chen2022beats}, and M2D~\cite{niizumi2023m2d}, learn low-level features by reconstructing individual masked patches during training, they incorporate irrelevant background information and are prone to semantic misalignment. Thus, they perform poorly in downstream tasks with limited labeled data, such as few-shot learning. This work investigates the integration of ID and MAEs for improving adaptation to downstream tasks.

\noindent \textbf{Instance Discrimination (ID)} methods, unlike MAM, align representations of different augmentations of an anchor example. Existing works utilize data augmentation techniques such as pitch/time shift~\cite{heggan2023mtslvr}, time mask/stretch~\cite{heggan2023mtslvr}, random crop and mixup~\cite{niizumi2021byol}, fade~\cite{heggan2023mtslvr}, mixed/white noise~\cite{heggan2023mtslvr}, and Gaussian noise~\cite{niizumi2021byol, saeed2021contrastive}. However, none of these methods apply ID for semantically aligning representations of pretrained MAEs. To the best of our knowledge, we are the first to propose an efficient CL strategy with unsupervised audio mixtures to semantically align pretrained MAE representations using only a small amount of unlabeled data. Our work is akin to recent work in the image domain~\cite{ViT2040} aiming to reduce unlabeled data requirements, and consequently, computational resources and training time by training Transformers with ID~\cite{lehner2023contrastive}.  
\section{Methodology}\label{sec:method}
\begin{figure}[t!]
    \centering
    \resizebox{\linewidth}{!}{
    \includegraphics[width=\columnwidth]{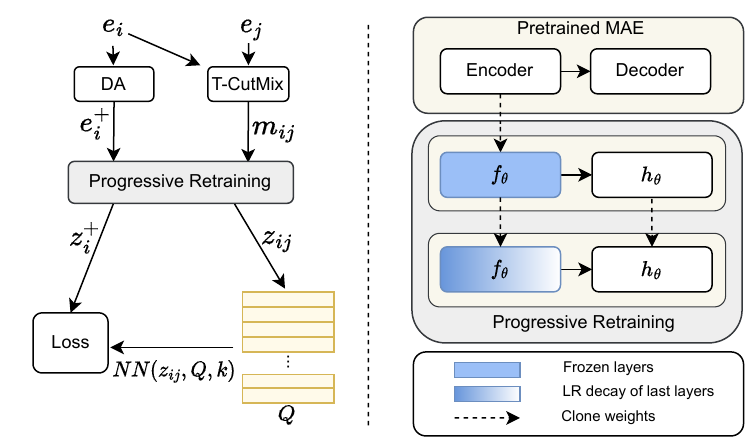}
    }
    \vspace{-0.9cm}
    \caption{\small \modelname overview. Left: T-CutMix contrastive tuning. Right: Progressive retraining of $f_\theta$ and $h_\theta$. DA: Data Augmentation.}
    \label{fig:architecture}
    %\vspace{-0.3cm}
\end{figure}

Given a pretrained MAE encoder $f_\theta$ and an unlabeled dataset $\mathcal{E}\!=\!\{(e_i,e_i^+)\}_{i=1}^N$, where $N$ is the number of total examples, $e_i \in \mathbb{R}^{T\times F}$ denotes the Filterbanks (fbanks)~\cite{fbank} of an audio sample $i$ with time $T$ and frequency $F$, and $e_i^+$ is a positive example of anchor $e_i$, constructed through data augmentation techniques, our objective is to improve downstream task performance with limited labeled data. To achieve this, we can employ ID methods to leverage unlabeled data for semantically aligning the representations of $f_\theta$ in the feature space. In practice, however, training with abundant unlabeled data is impractical for resource-constrained environments. Therefore, devising methods that extract maximal value from unlabeled data can augment the transferability and generalization capabilities of the learned representations while training with a small amount of unlabeled data. To this end, we introduce \modelname, which extends $f_\theta$ by incorporating a contrastive head $h_\theta$ and performs contrastive tuning on $h_\theta$ and the last layers of $f_\theta$. As the last layers capture more abstract and high-level features, contrastive tuning thus enhances the model's ability to utilize high-level semantics for downstream tasks. Fig.~\ref{fig:architecture} presents the overall architecture of \modelname.\\ 

\noindent \textbf{Contrastive Tuning Objective.} 
In terms of ID, we utilize the Nearest Neighbour Contrastive Learning (NNCLR) objective~\cite{dwibedi2021little}, an extension of SimCLR~\cite{chen2020simple} that utilizes a queue $Q$ for the nearest neighbor lookup of anchor examples. Specifically, given a batch $B\!=\!\{(z_i, z_i^+)\}_{i=1}^{|B|}$, where $z_i$ and $z_i^+$ are the feature representations of anchor $e_i$ and its positive example $e_i^+$, respectively, and $z_{j}$ denotes an example in $B$, the NNCLR loss function $\mathcal{L}_{CL}(z_i, z_i^+)$ is defined as follows: 
\begin{align}
\scalemath{0.85}{
\mathcal{L}_{CL}(z_i, z_i^+) \,{=}\, -\log \frac{\exp \big( NN(z_i, Q, k) \cdot z_i^+ / \tau \big) }
{ \sum_{(z_j, z_{j}^+) \in B }\exp \big( NN(z_i, Q, k) \cdot z_j^+ / \tau \big) }
}
\label{eq:nce}
\end{align}
Here, $\tau$ is a temperature hyperparameter and $NN(z_i, Q, k)$ denotes the top-$k$ nearest neighbors of $z_i$. The queue $Q$ is maintained similarly to MoCo~\cite{he2020momentum}. The use of NNCLR as the contrastive tuning objective is motivated by its ability to provide more semantic variations in the positive examples compared to other methods~\cite{he2020momentum,chen2020simple}.\\ 

\noindent \textbf{Progressive Retraining.} 
For contrastive tuning, we employ a progressive retraining strategy. Initially, we freeze the pretrained MAE encoder $f_\theta$ and train the head $h_\theta$ using the NNCLR objective. We then retrain the second half of $f_\theta$ along with the head $h_\theta$ using NNCLR. Partial retraining of $f_\theta$ is motivated by the success of partial finetuning, \ie tuning only the last layers~\cite{he2022masked}. Intuitively, the lower layers of the encoder are adept at generalization, capturing fundamental features that apply across various contexts, while retraining the upper layers enables high-level semantic alignment of features and invariance to subtle differences among examples.\\ 

\begin{figure}[t!]
    \centering
    \includegraphics[width=\columnwidth]{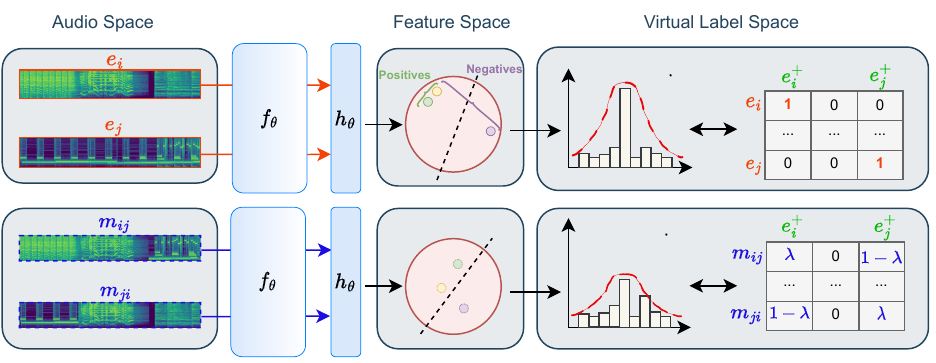}
    \vspace{-0.9cm}
    \caption{\small (Top row) When original audio samples $e_i$, $e_j$ are passed through $f_\theta$ and $h_\theta$, the positive pair is close to each other, and the negative pair lies far in the feature space, resulting in a sharp decision boundary in the virtual label space. (Bottom row) \modelname creates mixed audio samples $m_{ij}$ and $m_{ji}$ in the input space and uses a softened distance function in the virtual label space, resulting in a smoother decision boundary.}
    \label{fig:overview_figure}
    \vspace{-0.3cm}
\end{figure}
\noindent \textbf{Unsupervised T-Cutmix.}
Traditional audio augmentation methods typically construct a positive example $e_i^+$ by manipulating the audio sample $e_i$ solely in the input space. These techniques include pitch/time shift~\cite{heggan2023mtslvr}, time mask/stretch~\cite{heggan2023mtslvr}, noise~\cite{niizumi2021byol, saeed2021contrastive,heggan2023mtslvr}, random crop and mixup~\cite{niizumi2021byol}, \etc However, as depicted in Fig.~\ref{fig:overview_figure}, manipulating examples only in the input space results in a sharp decision boundary and, consequently, necessitating a large amount of data to learn generalized representations in the feature space~\cite{zhang2017mixup, yun2019cutmix}.
In contrast, we introduce T-Cutmix, an unsupervised mixing technique tailored for audio that manipulates data in both the audio and virtual label space as illustrated in Fig.~\ref{fig:overview_figure}. Our approach is akin to recent advances in image-based unsupervised mixing strategies such as MixUp~\cite{zhang2017mixup} and 2D-CutMix~\cite{yun2019cutmix}, which combine label smoothing and self-supervised virtual label space regularization~\cite{lee2021imix, shen2022unmix, dosovitskiy2020image, ViT2040}. 
Specifically, we define $v_i$ as the virtual label of $e_i$ and $e_i^+$, where $v_i[i]\,{=}\,1$ signifies that $e_i^+$ is the positive example of $e_i$ and $v_i[k \neq i]\,{=}\,0$ indicates that all other audio samples in the batch $B$ are considered negative examples in relation to $e_i$ (Fig.~\ref{fig:overview_figure} Virtual Label Space).

\modelname creates an audio mixture $m_{ij}$ and its corresponding smoothed label $y_{ij}$ as follows: 
\begin{align}
    m_{ij} &= \mathcal{M} \odot e_i + (1 - \mathcal{M}) \odot e_j  \\
    y_{ij} &= \lambda v_i + (1 - \lambda) v_j,
\end{align}
where $\lambda$ is a mixing coefficient and $\mathcal{M}\in\{0, 1\}^{T\times F}$ is a binary mask that determines which regions of an audio sample $e_i$ are replaced with corresponding regions from $e_j$, \ie how much information from each sample contributes to the mixture (Audio Space in Fig.~\ref{fig:overview_figure}).
To generate $\mathcal{M}$, a bounding box $BB = (s_t, 0, w_t, F)$ is sampled, indicating that $BB$ in $e_i$ is
replaced with the patch cropped from $BB$ of $e_j$. 
Here, $w_t = T\sqrt{1-\lambda}$ denoting the length of $BB$ in the time dimension. The starting time coordinate $s_t$ is uniformly sampled as $s_t\sim \text{Uniform}(0, T)$. For the starting coordinate and the length in the frequency dimension of $BB$, we keep the values fixed at $0$ and $F$, respectively. Similar to~\cite{lee2021imix, shen2022unmix}, $\lambda$ is sampled from a beta distribution $\mathcal{B}(\alpha, \alpha)$, where $\alpha$ is a hyperparameter controlling the size of $BB$. $\mathcal{M}$ is computed by filling the bounding box region with 0 and the rest with 1.
Finally, the loss function for audio mixture $m_{ij}$ is defined as
\begin{align}
\mathcal{L}(z_{ij}, y_{ij}) \,{=}\, -\sum\nolimits_{l=1}^{|B|} y_{ij}[l] \cdot \mathcal{L}_{CL}(z_{ij}, z_l^+),
\label{eq:nce_withvirtual_label}
\end{align}
where $z_{ij}$ is the representation of the mixed example $m_{ij}$, $y_{ij}[l]$ the $l$-th element of the smoothed virtual label $y_{ij}$, and $z_l^+$ the representation of the positive example in the batch.

\section{Experiments}\label{sec:experiments}
We evaluate \modelname on few-shot learning (FS) and finetuning (FT) downstream tasks across six benchmarks.

\setlength{\tabcolsep}{1.5pt}
\begin{table}[t]
  \caption{\small Dataset details for each evaluation setting. SSL: Self-supervised training, FT: Finetuning, FS: Few-shot learning}
  \centering
   \resizebox{\columnwidth}{!}{%
  \begin{tabular}{llccc}
    \toprule
    Dataset & Purpose & \# Classes & \# Samples & Audio Length \\
    \midrule
    AudioSet-20K \cite{gemmeke2017audio} & SSL \& FT & 527 & 20,550 & 10s \\
    ESC-50  \cite{piczak2015esc}     & FT \& FS         & 50   & 2,000      & 5s\\
    VoxCeleb1  \cite{nagrani2017voxceleb} & FT \& FS              & 1,251 & 153,516 & 3s - 180s\\
    SCv2 \cite{warden2018speech}   & FT \& FS       & 35 & 105,829     & 1s\\
    NSynth \cite{engel2017neural}   & FS       & 1,006 & 305,978     & 4s \\
    Kaggle18 \cite{fonseca2018general}  & FS                 & 41   & 11,073   & 0.3s - 30s \\
    
    \bottomrule
  \end{tabular}}
  \label{table:datasets}
\end{table}

\noindent\textbf{Baselines and Datasets.} We compare \modelname with other self-supervised MAEs trained with MAM. For all baselines, we use the publicly available pretrained checkpoints. 
The datasets used in the self-supervised training and the downstream tasks are detailed in Table~\ref{table:datasets}. For finetuning (FT), we use the AudioMAE train/validation/test splits~\cite{huang2022masked}.

\noindent \textbf{Training details.}
The backbone $f_\theta$ is a $86$M-parameter ViT-Base architecture. First, we initialize $f_\theta$ with pretrained AudioMAE encoder weights~\cite{huang2022masked} and train $h_\theta$ for $40$ epochs with learning rate $10^{-4}$, batch size $512$, temperature $\tau\,{=}\,0.15$, and $k\,{=}\,1$ in top $k$-NN lookup. Next, we freeze the lower half layers and train the top half layers by applying a layer-wise learning rate decay with decay factor $0.65$, learning rate $10^{-4}$ for $160$ epochs, and batch size $128$. Other hyperparameters are adopted from the NNCLR initialization and contrastive tuning steps in~\cite{lehner2023contrastive}. For few-shot learning, we follow~\cite{Ericsson2021HowTransfer}, and employ a nearest-centroid classifier on backbone extracted features. We report average accuracy ($95\%$ confidence interval) on $600$ randomly sampled few-shot episodes. For fine-tuning experiments, we follow the AudioMAE setup~\cite{huang2022masked}.

\subsection{Few-shot Learning}
\begin{table}[t!]

\centering
\caption{\small 5-way 1-shot performance using prototypical networks. Best performance is in \textbf{bold}. \#TP refers to \# of trainable parameters.}
\resizebox{\columnwidth}{!}{
\begin{tabular}{llccccc}
\toprule
\selectfont Method 
& {\selectfont \#TP}
& {\selectfont ESC-50}
& \selectfont VoxCe1eb1
& \selectfont NSynth & \selectfont SCv2 & \selectfont Kaggle18 \\
\midrule
MAE-AST~\cite{baade2022mae}&  99M & 49.3{\footnotesize$\pm$0.9}	&   25.6{\footnotesize$\pm$0.5}	&    48.7{\footnotesize$\pm$0.9} &   26.6{\footnotesize$\pm$0.5} &   38.4{\footnotesize$\pm$0.8}\\
MaskSpec~\cite{chong2023masked}&  86M & 43.0{\footnotesize$\pm$0.7}	&   -	&    - &   21.1{\footnotesize$\pm$0.4} &   -\\
BEATs~\cite{Chen2022beats}&   90M & 48.6{\footnotesize$\pm$0.8}	&   25.9{\footnotesize$\pm$0.5}	&    68.8{\footnotesize$\pm$0.9} &   26.9{\footnotesize$\pm$0.5} &   35.0{\footnotesize$\pm$0.8}\\
M2D~\cite{niizumi2023m2d}&  86M &  53.3{\footnotesize$\pm$0.9}	&   28.4{\footnotesize$\pm$0.6}	&    43.8{\footnotesize$\pm$0.9} &   30.1{\footnotesize$\pm$0.5} &   37.8{\footnotesize$\pm$0.8}\\
AudioMAE~\cite{huang2022masked} & 86M & 61.1{\footnotesize$\pm$0.9} 	&   28.9{\footnotesize$\pm$0.5} &    70.6{\footnotesize$\pm$0.9} &   \textbf{30.4{\footnotesize$\pm$0.5}} &   41.3{\footnotesize$\pm$0.8}\\
\midrule
\rowcolor{mygray} \modelname &  50M & \textbf{66.3{\footnotesize$\pm$0.9}}
&   \textbf{30.3{\footnotesize$\pm$0.5}}
&   \textbf{75.9{\footnotesize$\pm$0.9}}
&   29.6{\footnotesize$\pm$0.5} 
&   \textbf{43.6{\footnotesize$\pm$0.8}}
\\
\bottomrule
\end{tabular}
}
\label{tab:few_shot_results}
\vspace{-0.3cm}
\end{table}

Table~\ref{tab:few_shot_results} presents the 5-way 1-shot (5 classes, each with 1 example) comparison of \modelname against baselines using prototypical networks.
\modelname outperforms the best baseline by $4.90\%\textup{--}7.44\%$ in all datasets except SCv2. 
\begin{table}[t!]
\centering
\caption{\small 5-way 1-shot performance comparison among \modelname variants: No Mixing, MixUp + LS. Best performance in \textbf{bold}.}
\resizebox{\columnwidth}{!}{
\begin{tabular}{lccccc}
\toprule
\selectfont Method 
& {\selectfont ESC-50}
& \selectfont VoxCeleb1
& \selectfont NSynth & \selectfont SCv2 & \selectfont Kaggle18 \\
\midrule
No Mixing &   48.7{\footnotesize$\pm$0.8} 	&   23.8{\footnotesize$\pm$0.4} &    73.2{\footnotesize$\pm$0.9} &   \textbf{29.9{\footnotesize$\pm$0.5}} &   37.2{\footnotesize$\pm$0.8}\\
MixUp + LS &   62.6{\footnotesize$\pm$0.9} &   29.1{\footnotesize$\pm$0.5}	&   73.9{\footnotesize$\pm$0.9}	&   29.6{\footnotesize$\pm$0.5} &   41.8{\footnotesize$\pm$0.8}\\
\midrule
\rowcolor{mygray} \modelname &   \textbf{66.3{\footnotesize$\pm$0.9}}
&   \textbf{30.3{\footnotesize$\pm$0.5}}	
&   \textbf{75.9{\footnotesize$\pm$0.9}}
&   29.6{\footnotesize$\pm$0.5} 
&   \textbf{43.6{\footnotesize$\pm$1.8}}
\\
\bottomrule
\end{tabular}
}
\label{tab:ablation_with_mixup}
\vspace{-0.3cm}
\end{table}

\begin{figure}[t!]
    \centering
    \resizebox{.9\linewidth}{!}{
    \subfigure[{\textbf{N-way 1-shot}}]{\resizebox{.48\linewidth}{!}{\resizebox{\columnwidth}{!}{
\begin{tikzpicture}[every node/.append style={font=\huge}]
\begin{axis}[
    at={(0,0)},
    ymin=6,
    ymax=31,
    xmin=4,
    xmax=30,
    minor tick num =5,
    minor tick style={draw=none},
    minor grid style={thin,color=black!10},
    major grid style={thin,color=black!10},
	xlabel=N-way Used,
	ylabel=Accuracy(\%),
	width=12cm,height=7cm,
    tick align=outside,
    axis x line*=middle,
    axis y line*=none,
    ylabel style={align=center},
    xtick={1,5,10,15,20,25},
    legend style={nodes=right,draw=none},
    legend pos= north east,
    legend columns=1,
    every axis plot/.append style={ultra thick}
    ]
% AudioMAE
\addplot[color=red,mark=x] coordinates {
	(5, 28.89)
	(10, 18.07)
	(15, 13.76)
	(20, 11.66)
	(25, 10.01)
	(30, 9.05)
};

% BEATs
\addplot[color=blue,mark=o] coordinates {
	(5, 25.93)
	(10, 15.65)
	(15, 11.91)
	(20, 9.82)
	(25, 8.62)
    (30, 7.78)
	
};

% T-Cutmix
\addplot[color=black,mark=square] coordinates {
	(5, 30.31)
	(10, 19.92)
	(15, 15.66)
	(20, 13.62)
	(25, 12.02)
	(30, 11.00)
};

\legend{AudioMAE, BEATs, \modelname}
\end{axis}

\end{tikzpicture}
}} }
    \subfigure[{\textbf{5-way k-shot}}]{\resizebox{.48\linewidth}{!}{\resizebox{\columnwidth}{!}{

\begin{tikzpicture}[every node/.append style={font=\huge}]
\begin{axis}[
	at={(0,0)},
    ymin=23,
    ymax=51,
    xmin=1,
    xmax=26,
    minor tick num =5,
    minor tick style={draw=none},
    minor grid style={thin,color=black!10},
    major grid style={thin,color=black!10},
	xlabel=K-shot Used,
	ylabel=Accuracy(\%),
	width=12cm,height=7cm,
    tick align=outside,
    axis x line*=middle,
    axis y line*=none,
    ylabel style={align=center},
    xtick={1,5,10,15,20,25},
    legend style={nodes=right},
    legend pos= south east,
    legend columns=1,
    every axis plot/.append style={ultra thick}
    ]

% AudioMAE
\addplot[color=red,mark=x] coordinates {
	(1, 28.89)
	(5, 38.78)
	(10, 43.15)
	(15, 45.22)
	(20, 46.76)
	(25, 47.49)
};

% BEATs
\addplot[color=blue,mark=o] coordinates {
	(1, 25.93)
	(5, 30.49)
	(10, 33.10)
	(15, 33.76)
	(20, 34.81)
	(25, 36.19)
};

% T-Cutmix
\addplot[color=black,mark=square] coordinates {
	(1, 30.31)
	(5, 40.54)
	(10, 44.28)
	(15, 45.99)
	(20, 47.58)
	(25, 48.72)
};
\end{axis}

\end{tikzpicture}
}} }
    }
    \vspace{-0.4cm}
    \caption{\small N-way k-shot performance comparison on VoxCeleb1.}
    \label{fig:n_way_k_shot}
\end{figure}
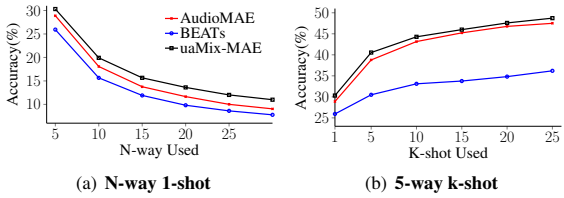

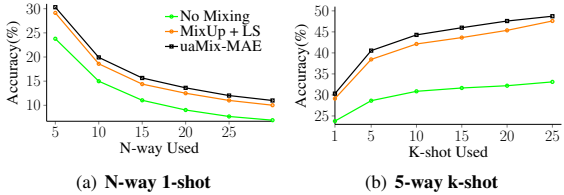
\begin{figure}[t!]
    \centering
\resizebox{.9\linewidth}{!}{

    \subfigure[{\textbf{N-way 1-shot}}]{\resizebox{.48\linewidth}{!}{\resizebox{\columnwidth}{!}{
\begin{tikzpicture}[every node/.append style={font=\huge}]
\begin{axis}[
    at={(0,0)},
    ymin=6.7,
    ymax=31,
    xmin=4,
    xmax=30,
    minor tick num =5,
    minor tick style={draw=none},
    minor grid style={thin,color=black!10},
    major grid style={thin,color=black!10},
	xlabel=N-way Used,
	ylabel=Accuracy(\%),
	width=12cm,height=7cm,
    tick align=outside,
    axis x line*=middle,
    axis y line*=none,
    ylabel style={align=center},
    xtick={1,5,10,15,20,25},
    legend style={nodes=right, draw=none},
    legend pos= north east,
    legend columns=1,
    every axis plot/.append style={ultra thick}
    ]
% No Mixing Strategy
\addplot[color=green,mark=o] coordinates {
	(5, 23.78)
	(10, 14.99)
	(15, 11.04)
	(20, 9.03)
	(25, 7.69)
    (30, 6.92)
	
};

% Mixup
\addplot[color=orange,mark=o] coordinates {
	(5, 29.13)
	(10, 18.59)
	(15, 14.39)
	(20, 12.50)
	(25, 11.01)
    (30, 10.02)
	
};

% T-Cutmix
\addplot[color=black,mark=square] coordinates {
	(5, 30.31)
	(10, 19.92)
	(15, 15.66)
	(20, 13.62)
	(25, 12.02)
	(30, 11.00)
};

\legend{No Mixing, MixUp + LS, \modelname}
\end{axis}

\end{tikzpicture}
}} }
    \subfigure[{\textbf{5-way k-shot}}]{\resizebox{.48\linewidth}{!}{\resizebox{\columnwidth}{!}{

\begin{tikzpicture}[every node/.append style={font=\huge}]
\begin{axis}[
	at={(0,0)},
    ymin=23,
    ymax=51,
    xmin=1,
    xmax=26,
    minor tick num =5,
    minor tick style={draw=none},
    minor grid style={thin,color=black!10},
    major grid style={thin,color=black!10},
	xlabel=K-shot Used,
	ylabel=Accuracy(\%),
	width=12cm,height=7cm,
    tick align=outside,
    axis x line*=middle,
    axis y line*=none,
    ylabel style={align=center},
    xtick={1,5,10,15,20,25},
    legend style={nodes=right},
    legend pos= south east,
    legend columns=1,
    every axis plot/.append style={ultra thick}
    ]

% No Mixing Strategy
\addplot[color=green,mark=o] coordinates {
	(1, 23.78)
	(5, 28.63)
	(10, 30.86)
	(15, 31.65)
	(20, 32.19)
	(25, 33.10)
};

% Mixup
\addplot[color=orange,mark=o] coordinates {
	(1, 29.13)
	(5, 38.45)
	(10, 42.12)
	(15, 43.64)
	(20, 45.36)
	(25, 47.59)
};

% T-Cutmix
\addplot[color=black,mark=square] coordinates {
	(1, 30.31)
	(5, 40.54)
	(10, 44.28)
	(15, 45.99)
	(20, 47.58)
	(25, 48.72)
};

\end{axis}

\end{tikzpicture}
}} }
}
\vspace{-0.4cm}
    \caption{\small Few-shot performance comparison on Voxceleb1 among \modelname variants: No Mixing and MixUp + LS.}
    \label{fig:ablation_graph_adding_mixup}
    \vspace{-0.4cm}
\end{figure}
 
\subsection{Few-shot Ablation Studies} 
\textbf{Varying N and K.} We vary N and K in the N-way K-shot experiments, where N is the class number, and K the number of examples per class.
Fig.~\ref{fig:n_way_k_shot} shows \modelname consistently outperforms baselines across different values of N and K.

\noindent\textbf{T-CutMix Importance.} We perform an ablation study considering the following variations: 1) No Mixing \ie employing no unsupervised mixing in CL, and 2) MixUp + LS \ie utilizing MixUp with label smoothing for unsupervised mixing. Results in Table~\ref{tab:ablation_with_mixup} and Fig.~\ref{fig:ablation_graph_adding_mixup} show substantial improvements compared to both variants across all scenarios.

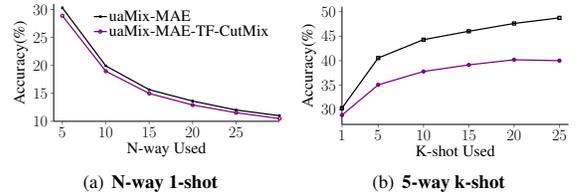
\begin{figure}[t!]
\centering
\resizebox{.9\linewidth}{!}{
    \subfigure[{\textbf{N-way 1-shot}}]{\resizebox{.48\linewidth}{!}{\resizebox{\columnwidth}{!}{
\begin{tikzpicture}[every node/.append style={font=\huge}]
\begin{axis}[
    at={(0,0)},
    ymin=10,
    ymax=31,
    xmin=4,
    xmax=30,
    minor tick num =5,
    minor tick style={draw=none},
    minor grid style={thin,color=black!10},
    major grid style={thin,color=black!10},
    xlabel=N-way Used,
    ylabel=Accuracy(\%),
    width=12cm,height=7cm,
    tick align=outside,
    axis x line*=middle,
    axis y line*=none,
    ylabel style={align=center},
    xtick={1,5,10,15,20,25},
    legend style={nodes=right,draw=none},
    legend pos= north east,
    legend columns=1,
    every axis plot/.append style={ultra thick}
    ]
% T-Cutmix
\addplot[color=black,mark=x] coordinates {
	(5, 30.31)
	(10, 19.92)
	(15, 15.66)
	(20, 13.62)
	(25, 12.02)
	(30, 11.00)
};

% TF-Cutmix
\addplot[color=violet,mark=o] coordinates {
	(5, 28.89)
	(10, 18.97)
	(15, 14.98)
	(20, 12.92)
	(25, 11.54)
    (30, 10.50)
	
};

\legend{\modelname, \modelname-TF-CutMix}
\end{axis}

\end{tikzpicture}
}} }
    \subfigure[{\textbf{5-way k-shot}}]{\resizebox{.48\linewidth}{!}{\resizebox{\columnwidth}{!}{

\begin{tikzpicture}[every node/.append style={font=\huge}]
\begin{axis}[
	at={(0,0)},
    ymin=27,
    ymax=51,
    xmin=1,
    xmax=26,
    minor tick num =5,
    minor tick style={draw=none},
    minor grid style={thin,color=black!10},
    major grid style={thin,color=black!10},
	xlabel=K-shot Used,
	ylabel=Accuracy(\%),
	width=12cm,height=7cm,
    tick align=outside,
    axis x line*=middle,
    axis y line*=none,
    ylabel style={align=center},
    xtick={1,5,10,15,20,25},
    legend style={nodes=right},
    legend pos= south east,
    legend columns=1,
    every axis plot/.append style={ultra thick}
    ]

% T-Cutmix
\addplot[color=black,mark=square] coordinates {
	(1, 30.31)
	(5, 40.54)
	(10, 44.28)
	(15, 45.99)
	(20, 47.58)
	(25, 48.72)
};

% TF_Cutmix 
\addplot[color=violet,mark=*] coordinates {
	(1, 28.89)
	(5, 35.07)
	(10, 37.79)
	(15, 39.14)
	(20, 40.18)
	(25, 40.00)
};

\end{axis}

\end{tikzpicture}
}} }

}
\vspace{-0.4cm}
    \caption{\small Few-shot performance comparison between \modelname and \modelname-TF-CutMix on VoxCeleb1.}
    \label{fig:cutmix_ablation}
\end{figure}

\noindent\textbf{TF-CutMix.} To illustrate the impact of applying CutMix exclusively in the time dimension, we conduct an ablation study introducing a variation, termed \modelname-TF-CutMix, that employs CutMix in both time (T) and frequency (F) dimensions. As depicted in Fig.~\ref{fig:cutmix_ablation}, \modelname with T-CutMix consistently outperforms \modelname with TF-CutMix.

\subsection{Fine-tuning}
\begin{table}[t!]

\centering
\caption{\small Fine-tuning performance on audio and speech classification
tasks. Best performance in \textbf{bold}. }
\resizebox{\columnwidth}{!}{
\begin{tabular}{llccccc}
\toprule
\selectfont Method 
& {\selectfont \#TP}
& {\selectfont AudioSet-20k}
& \selectfont ESC-50
& \selectfont VoxCeleb1
& \selectfont SCv2\\
\midrule
{BEATs~\cite{Chen2022beats}} & {90M} &   {36.0} 	&   {94.0} &   {-} & \textbf{98.3} \\
{MAE-AST~\cite{baade2022mae}} & {99M} &   {30.6} 	&   {90.0} &   {63.3} & {97.9} \\
MaskedSpec~\cite{chong2023masked} & 86M &   32.3 	&   89.6 &    - & 97.7 \\
AudioMAE~\cite{huang2022masked} & 86M &   36.7 	&   94.0 &    93.5 & 97.9 \\
\midrule
\rowcolor{mygray}\modelname & 86M &   \textbf{37.0} 
&   \textbf{94.1}
&   \textbf{93.6}
& {98.0}
\\
\bottomrule
\end{tabular}
}
\label{tab:fine_tune_results}
\vspace{-0.3cm}
\end{table}

Table~\ref{tab:fine_tune_results} presents a fine-tuning comparison with \modelname demonstrating comparable results {w.r.t.} other baselines across all datasets. 
Table~\ref{tab:few_shot_results} and~\ref{tab:fine_tune_results} indicate that \modelname achieves better performance in few-shot learning, \ie demonstrating superior generalization in the feature space while maintaining competitive results in fine-tuning.

\subsection{Qualitative Analysis}
\begin{figure}[t!]
    \centering
    \resizebox{0.95\linewidth}{!}{
    \includegraphics[width=\columnwidth]{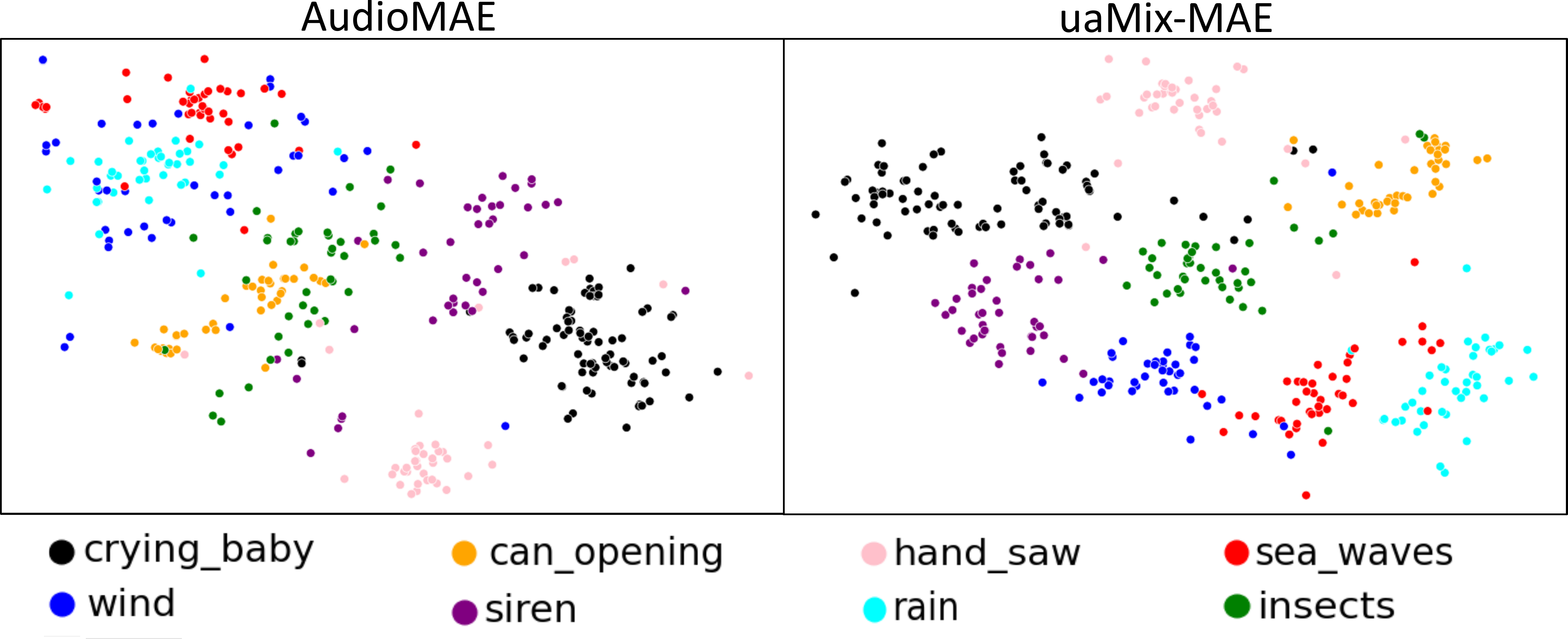}
    }
    \vspace{-0.4cm}
    \caption{\small t-SNE visualization of AudioMAE (left) and \modelname (right) features for eight ESC-50 classes.}
    \label{fig:tsne_visualization}
\vspace{-0.3cm}
\end{figure}
We compare the learned feature representations of the AudioMAE encoder and \modelname. The t-SNE visualization~\cite{van2008visualizing} for eight ESC-50 classes (Fig.~\ref{fig:tsne_visualization})
reveals that \modelname exhibits better intra-class clustering compared to AudioMAE. Specifically, \modelname representations form distinct and well-separated clusters for classes `{\color{red}see\_waves}', `{\color{blue}wind}', `{\color{violet}siren}', and `{\color[wave]{485}rain}' while the AudioMAE representations for these classes overlap with other classes.

\section{Conclusion}\label{sec:conclusion}
In this paper, we introduce \modelname, a contrastive tuning strategy employing unsupervised audio mixtures. To adapt to downstream tasks with limited labeled data, \modelname tunes a pretrained MAE encoder with a small amount of unlabeled data by mixing examples in both input and virtual label spaces. Experiments in few-shot settings demonstrate that \modelname outperforms existing masked audio models.


\newpage

%%%%%%%%% REFERENCES

\bibliographystyle{IEEEtraN}
{\small
\bibliography{egbib}
}

\end{document}